\documentclass[conference]{IEEEtran}
\IEEEoverridecommandlockouts
\usepackage{cite}
\usepackage{amsmath,amssymb,amsfonts}
\usepackage{algorithmicx}
\usepackage{algorithm}
\usepackage{algpseudocode}
\usepackage{graphicx}
\usepackage{textcomp}
\usepackage{xcolor}
\usepackage{soul}
\usepackage{subfig}
\usepackage{stfloats}

\def\BibTeX{{\rm B\kern-.05em{\sc i\kern-.025em b}\kern-.08em
    T\kern-.1667em\lower.7ex\hbox{E}\kern-.125emX}}
\begin{document}

\title{A Case Study in Responsible AI-Assisted Video Solutions: Multi-Metric Behavioral Insights in a Public Market Setting}


\author{\IEEEauthorblockN{1\textsuperscript{st} Mehrnoush Fereydouni}
\IEEEauthorblockA{\textit{BioInformatics} \\
\textit{UNC Charlotte}\\
Charlotte, USA \\
mfereydo@charlotte.edu}
\and
\IEEEauthorblockN{2\textsuperscript{nd} Eka Ebong}
\IEEEauthorblockA{\textit{Electrical and Computer Engineering} \\
\textit{University of North Carolina at Charlotte (UNC Charlotte)}\\
Charlotte, USA \\
eebong1@charlotte.edu}
\and

\IEEEauthorblockN{3\textsuperscript{rd} Sahar Maleki}
\IEEEauthorblockA{\textit{Belk College of Business} \\
\textit{UNC Charlotte}\\
Charlotte, USA \\
smaleki1@charlotte.edu}
\and
\IEEEauthorblockN{4\textsuperscript{th} Philip Otienoburu}
\IEEEauthorblockA{\textit{Community Engagement VP} \\
\textit{Charlotte Center City Partners}\\
Charlotte, USA \\
potienoburu@charlottecentercity.org}
\and
\IEEEauthorblockN{5\textsuperscript{th} Babak Rahimi~Ardabili}
\IEEEauthorblockA{\textit{Electrical and Computer Engineering} \\
\textit{UNC Charlotte}\\
Charlotte, USA \\
brahimia@charlotte.edu}
\and
\IEEEauthorblockN{6\textsuperscript{th} Hamed Tabkhi}
\IEEEauthorblockA{\textit{Electrical and Computer Engineering} \\
\textit{UNC Charlotte}\\
Charlotte, USA \\
htabkhiv@charlotte.edu}
}

\maketitle

\begin{abstract}

Despite recent advances in Computer Vision and Artificial Intelligence (AI), AI-assisted video solutions have struggled to penetrate real-world urban environments due to significant concerns regarding privacy, ethical risks, and technical challenges like bias and explainability. This work addresses these barriers through a case study in a city-center public market, demonstrating a pathway for the responsible deployment of AI in community spaces. By adopting a user-centric methodology that prioritizes public trust and privacy safeguards, we show that detailed, operationally relevant behavioral insights can be derived from abstract data representations without compromising ethical standards. The study focuses on generating Multi-Metric Behavioral Insights through the extraction of three complementary signals: customer directional flow, dwell duration, and movement patterns. Utilizing human pose detection and complex behavioral analysis—processed through geometric normalization and motion modeling—the system remains robust under tracking fragmentation and occlusion. Data collected over $18$ days, spanning routine operations and a festival window from May $2$--$4$, reveals a consistently right-skewed dwell-time behavior. While most visits last approximately $3$--$4$ minutes, peak activity periods increase the mean to roughly $22$ minutes. Furthermore, movement analysis indicates uneven circulation, with over $60\%$ of traffic concentrated in approximately $30\%$ of the venue space. By mapping popular thoroughfares and high-traffic storefronts, this case study provides venue managers and business owners with objective, measurable information to optimize foot traffic. Ultimately, these results demonstrate that AI-enabled video solutions can be successfully integrated into urban environments to provide high-fidelity spatial analytics while maintaining strict adherence to privacy and social responsibility.

\end{abstract}


\section{Introduction}

Video-based analytics are increasingly vital for supporting decision-making in urban environments, retail spaces, and public venues, enabling data-driven operational planning and safety-aware infrastructure design \cite{b20}. However, the majority of existing systems rely on pixel-level video and appearance-based representations, which often involve identity tracking or facial recognition. These methods raise substantial privacy concerns and frequently conflict with emerging governance and regulatory principles for AI deployment in public spaces. Beyond ethical and legal risks, reliance on raw imagery introduces significant deployment barriers, erodes public trust, and constrains the long-term adoption of AI technologies \cite{b21, b22, b37}.To overcome these barriers, this work presents a case study in responsible AI deployment, utilizing a framework that prioritizes privacy-preserving observations through abstract representations of human activity. We build on an existing AI-enabled video analytics pipeline rather than proposing a new end-to-end system. Specifically, our contribution is a multi-metric behavioral insight layer implemented on top of the established Ancilia platform \cite{b1}, which provides the underlying sensing, detection, and tracking infrastructure. This design choice allows us to focus on higher-level analytics without modifying the core edge pipeline. By using pose-based or skeletal motion data, the system removes facial features, textures, and personal identifiers at the earliest stages of processing \cite{b1}. While providing strong privacy guarantees, such abstractions introduce technical challenges for spatial reasoning and tracking stability in crowded, occluded settings \cite{b23}. Extracting meaningful behavioral metrics from this sparse data requires specialized analytical strategies that integrate geometric normalization, motion modeling, and robust statistical learning.In this study, we extract three complementary behavioral data types from these privacy-preserving streams: directional flows, dwell times, and movement patterns. These metrics represent a shift from raw detection counts toward interpretable behavioral understanding, offering actionable insights for layout design, staffing strategies, and crowd management. Specifically, our contribution demonstrates how to maintain tracking consistency and estimate dwell times without appearance cues or persistent identities—addressing issues like track fragmentation and sensor noise through physics-informed motion modeling and robust statistical aggregation \cite{b24, b25, b26, b27}.The proposed system was deployed in a real-world public market in a major U.S. city, collecting data over an $18$-day period that included both routine operations and high-traffic event windows. The infrastructure leverages the Ancilia sensing pipeline \cite{b1} for pose-based detection, upon which we integrate a behavioral analytics layer. 

Our approach operates exclusively on abstract metadata, ensuring no privacy-sensitive information is reintroduced.Across the deployment, our results provide objective evidence of the system's utility. Dwell-time distributions consistently exhibited a right-skewed pattern, with most visits lasting approximately $3$--$4$ minutes (median $\approx 3.6$ minutes), while peak activity increased the mean to roughly $22$ minutes. During a festival period (May $2$--$4$), bidirectional flow increased substantially yet remained symmetric, indicating efficient throughput rather than crowd accumulation. Furthermore, movement patterns identified high-traffic thoroughfares and the relative effectiveness of individual storefronts in attracting foot traffic.

Our main contributions are:
\begin{itemize}
\item A case study demonstration of a responsible AI framework for extracting multi-metric behavioral data (flows, dwell times, and patterns) from abstract pose-based representations.
\item A privacy-preserving data pipeline that integrates motion modeling and statistical learning to provide high-fidelity insights without facial recognition or identity storage.
\item Empirical validation through a real-world public market deployment, illustrating how operationally relevant spatial insights can be derived while maintaining strict adherence to social and ethical standards.
\end{itemize}

\section{Related Work}

Prior work on behavioral analytics from video can be grouped into pixel-based approaches and privacy-preserving representations. Early systems relied on background subtraction and blob tracking but were brittle under illumination changes and dynamic backgrounds \cite{b19}. Recent methods combine deep learning-based detection with multi-object tracking to improve performance in crowded scenes \cite{b2}, typically integrating motion models with appearance embeddings to maintain identity consistency \cite{b3}. Despite these advances, identity switches and tracking fragmentation remain prevalent under heavy occlusion and long-term deployments \cite{b4}. Trajectory smoothing, re-identification, and multi-camera fusion partially mitigate these issues \cite{b5} but introduce additional computational overhead and rely on persistent appearance features.

Geometric reasoning over human motion has been widely studied using geometric transformations for estimating crowd density, spatial utilization, and pedestrian flows \cite{b6}. While enabling spatial interpretation of movement and interactions, most pipelines rely on pixel-level detections or dense features and remain sensitive to calibration errors and perspective distortion \cite{b7}, while posing privacy risks due to reliance on raw visual cues.

In response, privacy-preserving video analytics leverage top-view sensing, edge-based processing, and abstract human representations. Ceiling-mounted cameras reduce identity exposure and occlusion \cite{b8}, while edge analytics transmit only anonymized metadata \cite{b9}. Pose-based abstractions further remove facial and appearance information while preserving motion dynamics \cite{b10}. However, extracting stable behavioral metrics from pose data remains challenging due to tracking fragmentation, sparse spatial cues for geometric transformation, and reduced robustness under occlusion \cite{b11}, motivating hybrid strategies that combine physics-informed motion modeling with statistical learning.

Behavioral metrics derived from video have demonstrated practical value across retail and public domains. In retail, dwell time correlates with customer engagement \cite{b12}, spatial heatmaps inform layout design \cite{b13}, entrance/exit flows guide staffing and queue management \cite{b14}, and movement patterns reveal circulation paths and bottlenecks \cite{b15}. Beyond retail, crowd flow and occupancy analytics support pedestrian routing, congestion mitigation, and safety interventions in public venues and event spaces \cite{b16,b17,b18}. However, most existing solutions depend on pixel-level or identity-linked representations, limiting deployment in privacy-sensitive contexts. This gap motivates extracting high-value behavioral metrics from abstract, privacy-preserving pose-based data without adding computational burden to the underlying AI pipeline.

\section{System Setup and Data Collection} 

\begin{figure*}[]
\centering
    \includegraphics[width=0.95\textwidth, trim=60 0 30 50]{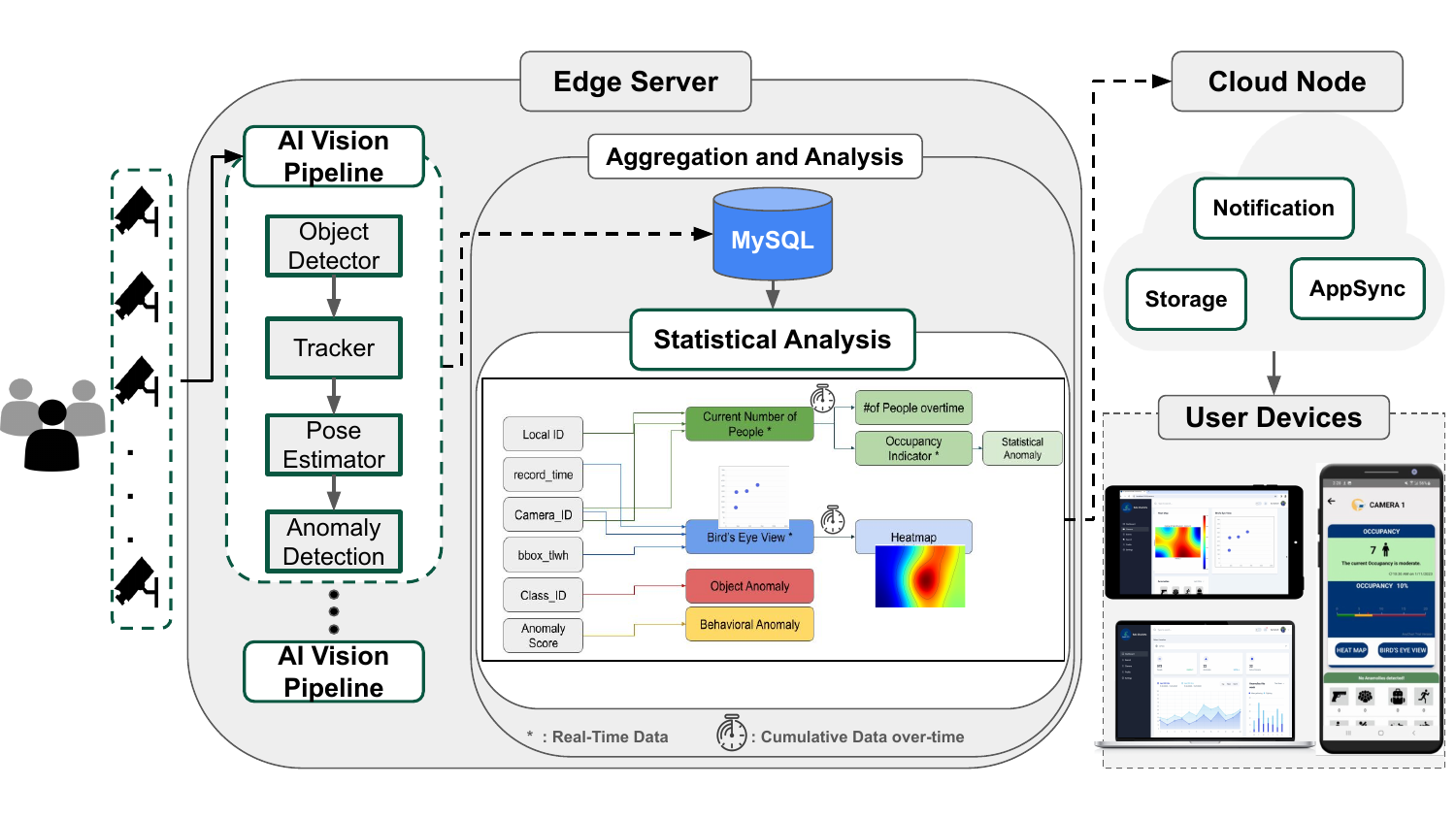}
\caption{End-to-end system architecture adapted from the Ancilia platform \cite{b1}. All analytical components are performed at the statistical analysis stage using metadata extracted from the AI vision pipeline.}
\label{fig:end2end}
\end{figure*}

This section describes the server-side methods for data collection, processing, and storage. As shown in Fig.~\ref{fig:end2end}, the underlying sensing and vision pipeline is adapted from the Ancilia system \cite{b1}, while our contribution focuses on the statistical analysis layer that derives multi-metric behavioral insights from pose-based metadata.
Live CCTV streams are processed in real time by an AI-enabled computer vision pipeline running on a dedicated on-site server. Video feeds are transmitted over Wi-Fi to the server, where machine learning models detect and track objects with a primary focus on human activity \cite{b1}. For each detected entity, the pipeline generates structured metadata, including object class identifiers, local tracking IDs, bounding box coordinates, and feature embeddings (for human detections). Each record is time-stamped and labeled with camera and batch identifiers before being archived in the server-side database for downstream analysis and visualization.

Data were collected over an 18-day period from April 25 to May 12, 2025, at a city-center public market in the United States. Data were not collected on May 8 because the system was undergoing an upgrade; as a result, no data are available for that day. The system runs on a dedicated server equipped with a 16-core CPU, 252~GB RAM, and four GPUs (24~GB VRAM each), supporting real-time inference and high-throughput processing. All video processing occurs on-site, and only anonymized, structured metadata are retained in the database for analysis.

The archived records are designed to be machine-readable rather than directly interpretable by end users. Stored entries consist solely of abstract identifiers, geometric descriptors, timestamps, and feature embeddings, without raw video frames or human-recognizable attributes. This design enables scalable downstream analytics and cross-stream aggregation while minimizing privacy risks and preventing direct human interpretation of individual-level information.

\section{Results} 

\subsection{Dwell Time}

\begin{figure}[h]
    \centering
    \includegraphics[width=\linewidth, trim=60 0 60 0, clip]{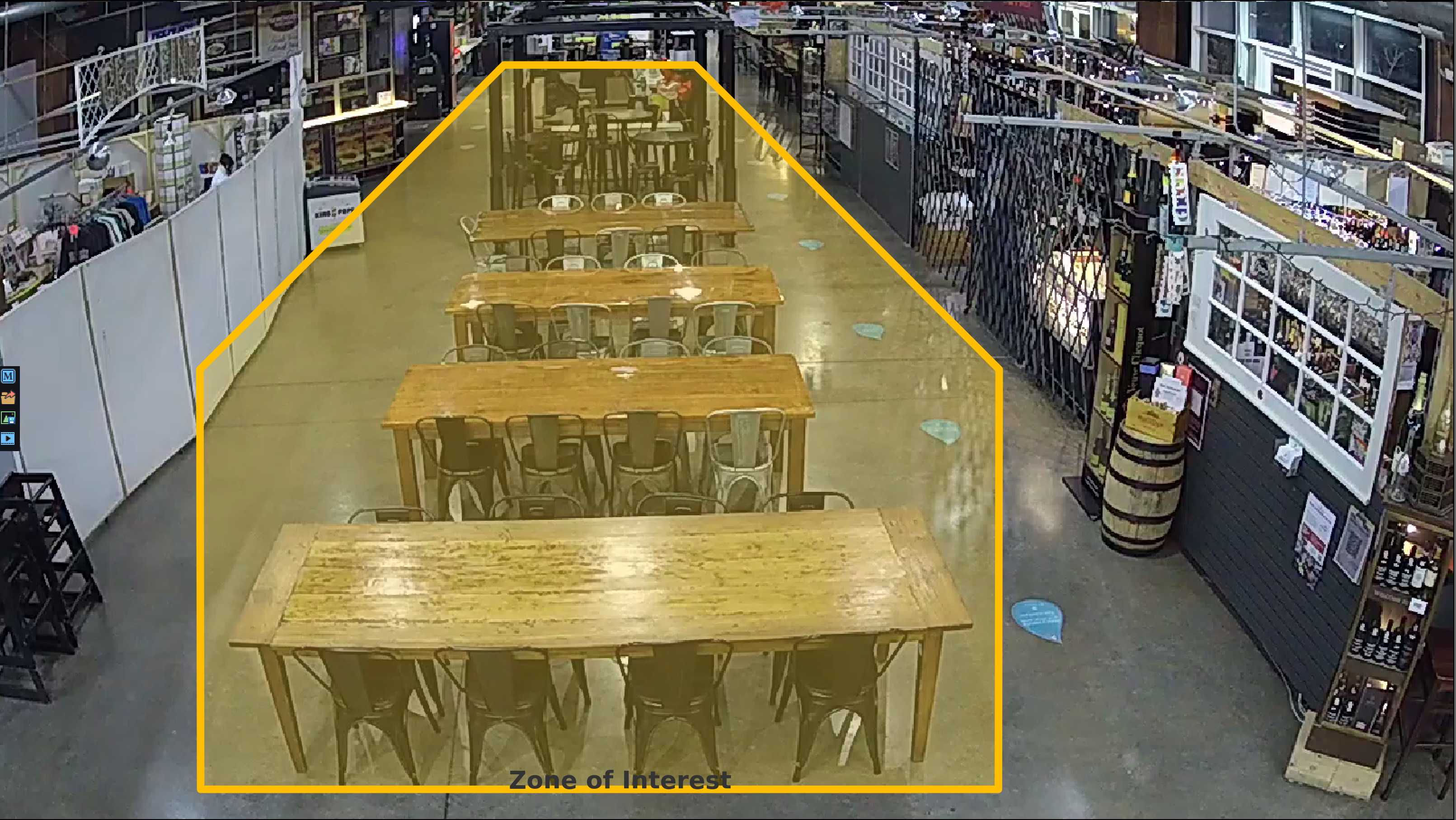}
    \caption{Seating ZOI for Camera~1 in pixel coordinates.}
    \label{fig:zoi_camera1}
\end{figure}

Movement trajectories alone do not fully capture spatial engagement, as entry counts ignore duration and cannot distinguish brief pass-throughs from sustained presence. This limitation is particularly relevant in privacy-preserving analytics, where identity information is unavailable. \emph{Dwell time} addresses this gap by measuring how long individuals remain within a region without reintroducing identity information. Dwell time was computed from raw tracking data collected by Camera~1, which directly views the primary seating area. Because tracking jitter can fragment stationary behavior, analysis was restricted to a manually annotated seating Zone of Interest (ZOI) defined in camera pixel coordinates. The ZOI was created as a binary mask and converted into a simplified polygon (Fig.~\ref{fig:zoi_camera1}). For each detection, the anchor point was defined as the bottom-center of the bounding box (Eq.~\eqref{eq:anchor_point}):

\begin{equation}
(c_x, c_y) = \left(x + \frac{w}{2},\, y + h\right),
\label{eq:anchor_point}
\end{equation}
A detection was retained if $(c_x,c_y)$ lay within the ZOI polygon using a point-in-polygon test.

\begin{figure}[h]
    \centering
    \includegraphics[width=\linewidth, trim=0 20 0 20, clip]{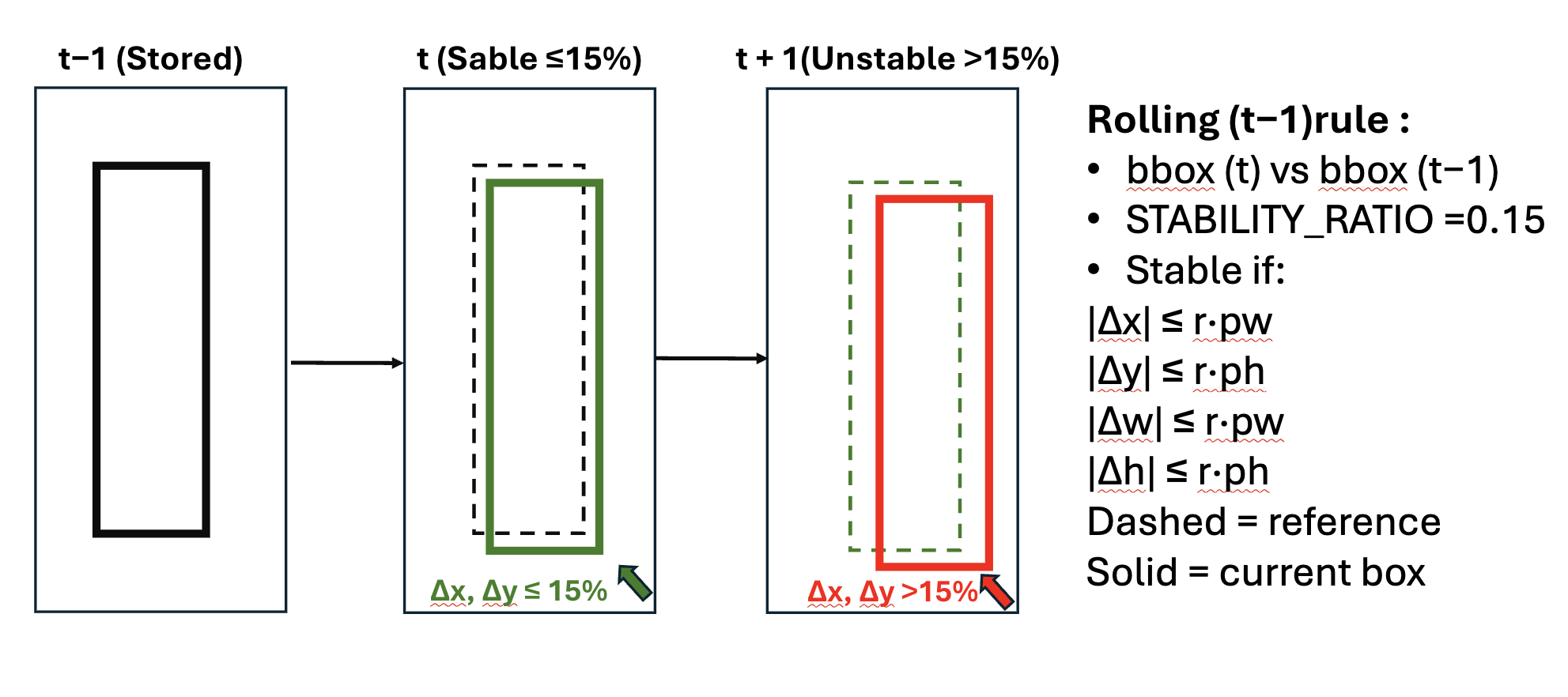}
    \caption{Rolling $(t{-}1)$ stability rule (15\% threshold).
}
    \label{fig:tminus1_rules}
\end{figure}

Stationarity was evaluated using a rolling $(t{-}1)$ comparison of bounding-box changes in $(x,y,w,h)$. Detections were considered stable when the maximum relative change was below 15\%, a value selected to balance suppression of tracking jitter with sensitivity to genuine movement. Fig.~\ref{fig:tminus1_rules} illustrates the $(t{-}1)$ rule. Visual inspection of overlaid bounding boxes confirmed that this threshold suppressed noise while preserving stationary behavior. Time-based constraints were applied. Stability was required for at least 2~seconds before initiating the dwell timer. Meaningful dwell events required a minimum duration of 60~seconds. A maximum dwell limit of 2~hours (7200~seconds) was enforced due to periodic system restarts and identifier resets, preventing artificial inflation.

\begin{table}[h]
\caption{Daily summary of individual dwell times within the ZOI for Camera~1 (seconds).}

\label{tab:dwell_daily_summary}
\centering
\begin{tabular}{lrrrr}
\hline
Date & $N$ & Mean (s) & Median (s) & SD (s) \\
\hline
2025-04-25 & 19  & 366.6  & 180.0 & 701.1 \\
2025-04-26 & 190 & 1045.6 & 154.0 & 1773.4 \\
2025-04-27 & 61  & 920.9  & 224.0 & 1400.6 \\
2025-04-28 & 22  & 572.2  & 180.0 & 1279.5 \\
2025-04-29 & 2   & 1159.0 & 1159.0 & 1516.0 \\
2025-04-30 & 32  & 827.2  & 182.5 & 1256.6 \\
2025-05-01 & 123 & 931.6  & 163.0 & 1730.5 \\
2025-05-02 & 138 & 1338.7 & 216.0 & 2178.2 \\
2025-05-03 & 47  & 638.5  & 112.0 & 1317.9 \\
2025-05-04 & 180 & 1528.3 & 234.0 & 2145.3 \\
2025-05-05 & 68  & 617.1  & 108.0 & 1399.9 \\
2025-05-06 & 66  & 1011.9 & 191.0 & 1664.3 \\
2025-05-07 & 71  & 1024.2 & 208.5 & 1626.9 \\
2025-05-09 & 98  & 960.5  & 145.0 & 1739.3 \\
2025-05-10 & 16  & 327.0  & 120.5 & 595.0 \\
2025-05-11 & 38  & 434.5  & 123.5 & 784.6 \\
2025-05-12 & 50  & 1040.3 & 155.0 & 1635.9 \\

\hline
\end{tabular}
\end{table}

\begin{figure}[h]
    \centering
    \includegraphics[width=0.85\linewidth]{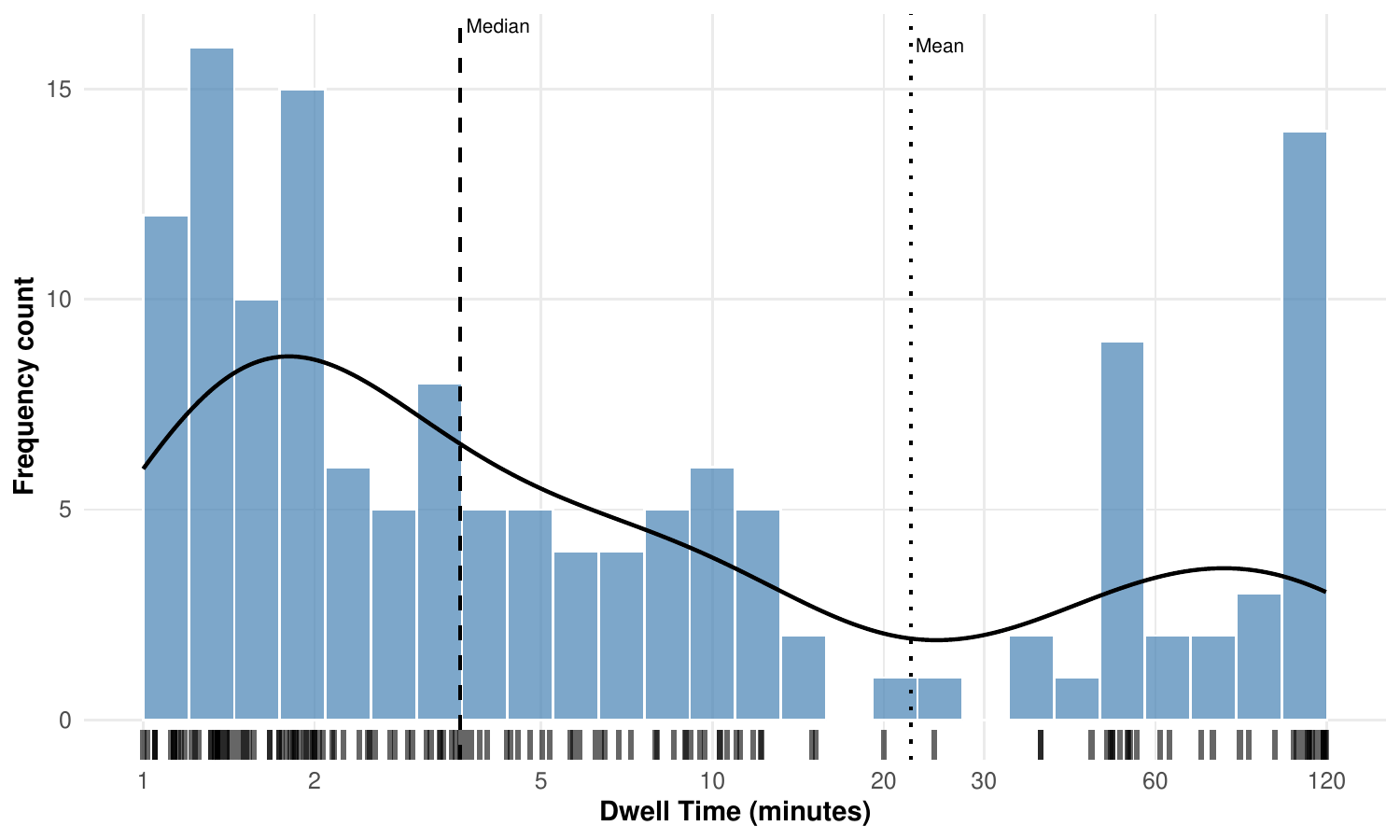}
    \caption{Distribution of dwell times within the Camera~1 ZOI on May~2,~2025.}
    \label{fig:dwell_may02_distribution}
\end{figure}
\begin{figure*}[t]
    \centering
    \includegraphics[width=0.8\linewidth]{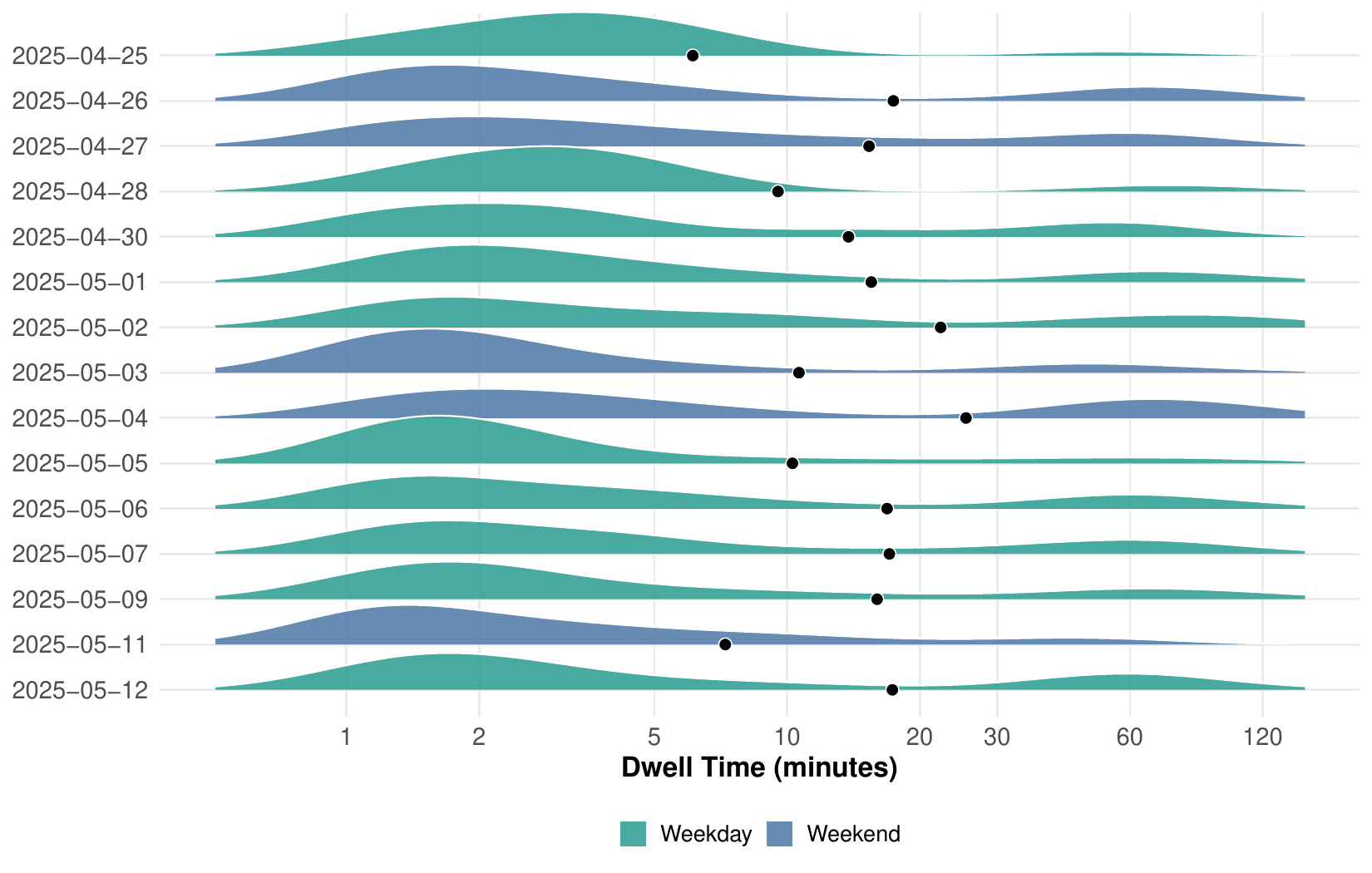}
    \caption{Daily dwell-time distributions by day, with black markers representing the mean dwell time.}
    \label{fig:ridge_by_day}
\end{figure*}

Table~\ref{tab:dwell_daily_summary} reports day-level summary statistics for the extracted individual dwell times within the same zone. April~29 and May~10 exhibit markedly lower observation counts (Table~\ref{tab:dwell_daily_summary}) and were excluded from subsequent visual analyses due to sparsity. While Table~\ref{tab:dwell_daily_summary} reports aggregate statistics, it does not reveal distributional structure. A representative high-activity day is therefore examined. Figure~\ref{fig:dwell_may02_distribution} shows a strongly right-skewed distribution on May~2,~2025. Most visitors remain seated for 3--4~minutes (median $\approx 3.6$~minutes), while a smaller subset exceeds 30~minutes. These long stays increase the mean to approximately 22~minutes, producing divergence between the median and mean and indicating mixed-use behavior. Figure~\ref{fig:ridge_by_day} extends the analysis across days. Distributions remain right-skewed, with short stays clustered around 3--4~minutes and a smaller fraction exceeding 30~minutes. Inter-day variability appears in distribution shape and mean dwell time. Days with heavier right tails exhibit right-shifted means, and weekend days show broader distributions with greater density at longer durations, indicating increased sustained engagement.

\subsection{Entry/exit counts}
\begin{figure*}[t]
    \centering
    \subfloat[Camera 3]{%
        \includegraphics[width=0.48\textwidth, trim=60 0 60 0, clip]{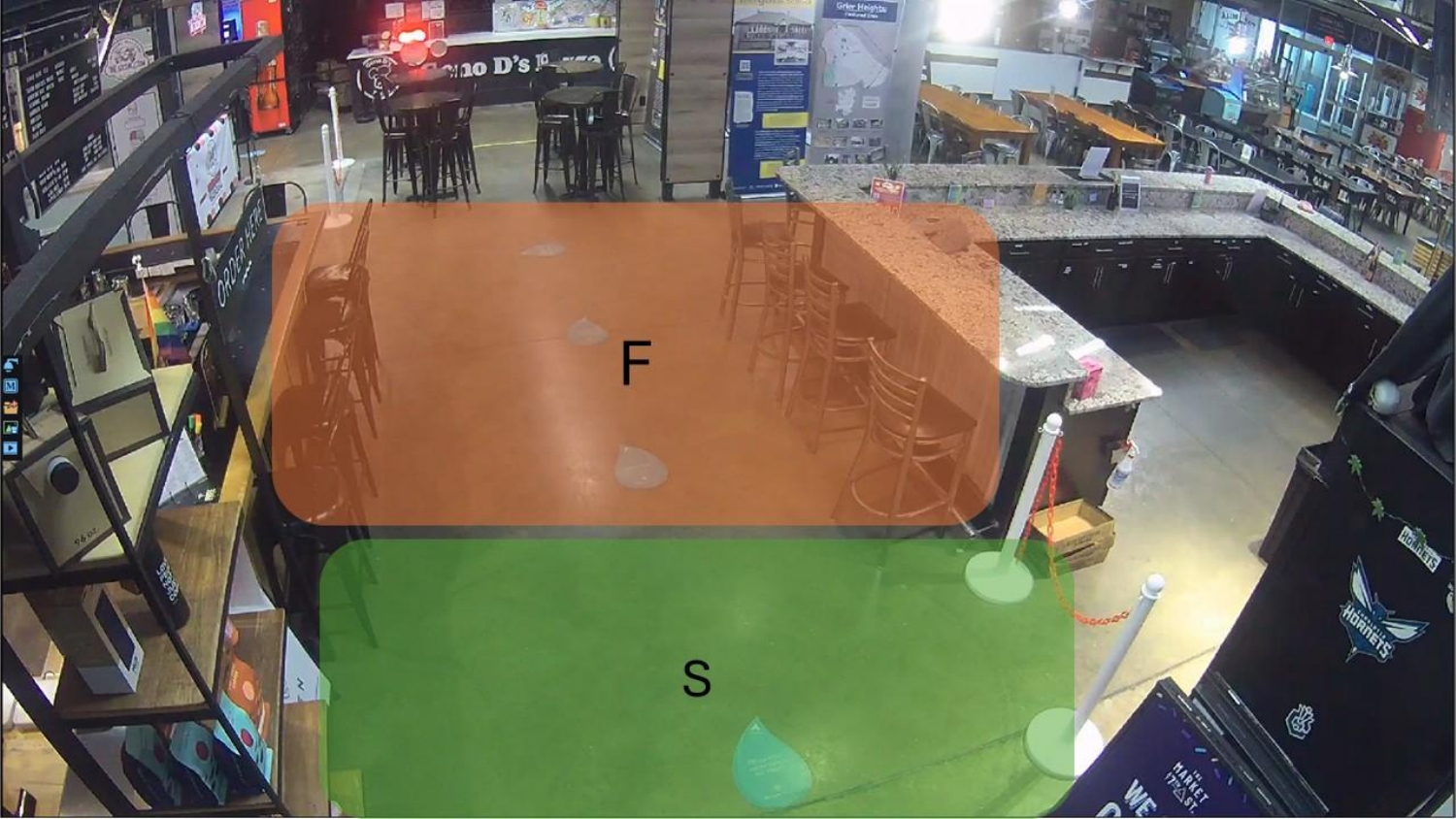}
    }
    \hfill
    \subfloat[Camera 2]{%
        \includegraphics[width=0.48\textwidth, trim=60 0 60 0, clip]{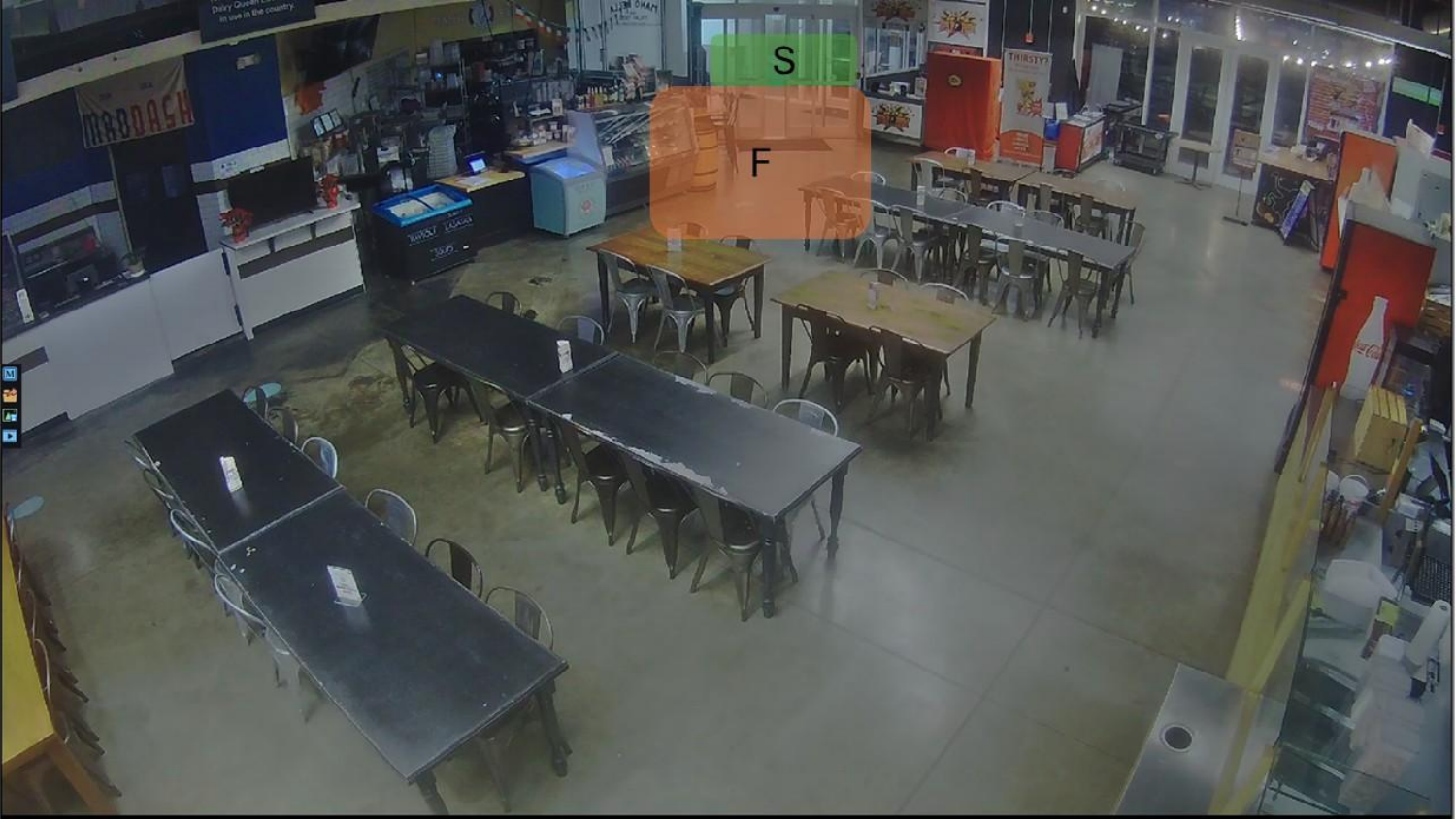}
    }
    \caption{Comparison of views from two camera locations.}
    \label{fig:camera_views}
\end{figure*}

\begin{figure*}
\centering
    \includegraphics[width=\textwidth, trim=70 60 50 0]{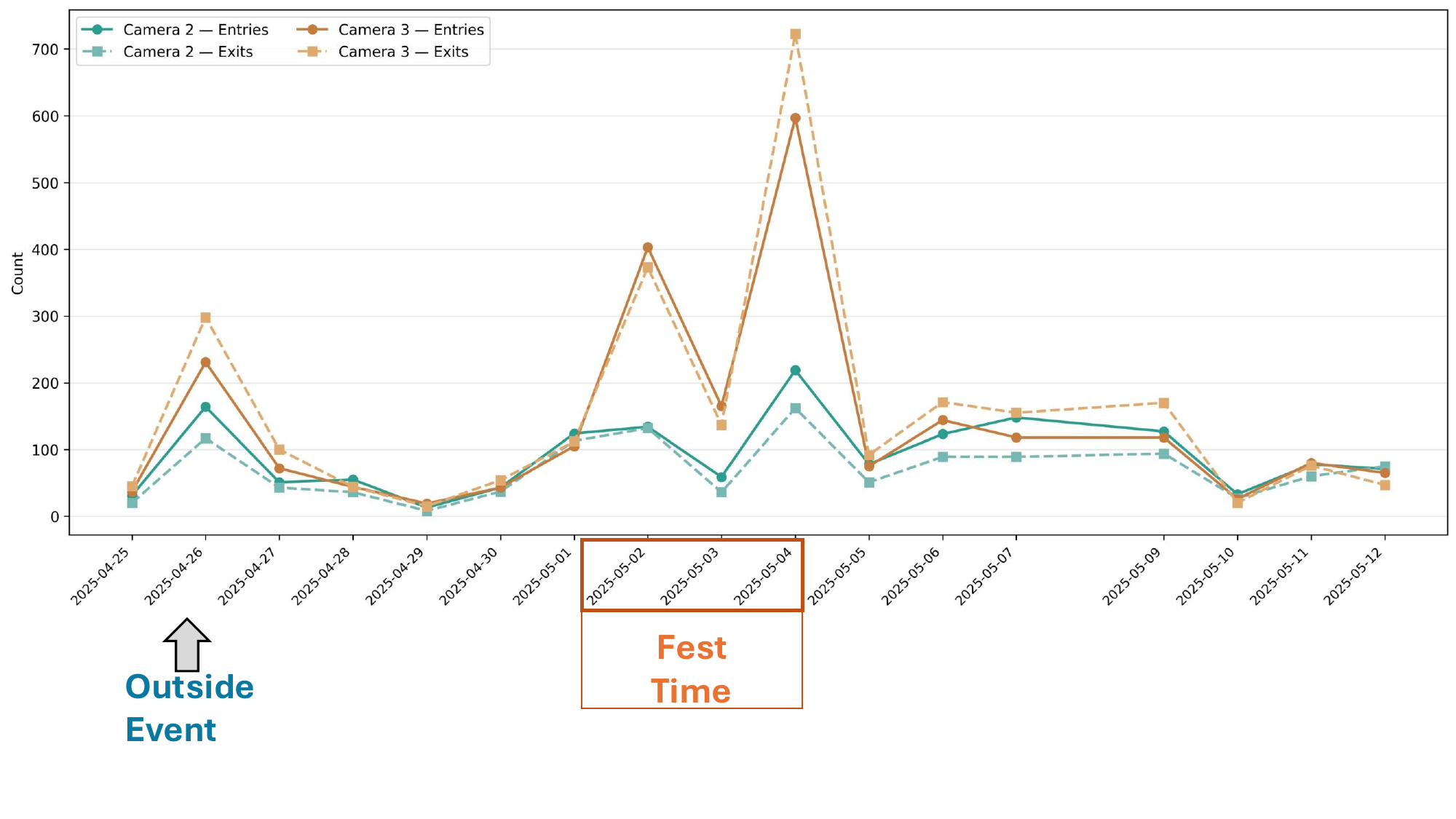}
\caption{Daily entry and exit counts for Cameras 2 and 3.}
\label{fig:daily_entry_exit}
\end{figure*}

To quantify directional movement, two spatial regions were manually defined within each camera’s field of view: a Start zone (S) and a Finish zone (F), as shown in Fig.~\ref{fig:camera_views}. These zones represent the logical flow boundary of the monitored space, where S corresponds to the entry side and F to the opposite traversal boundary. For each detection, the bounding box centroid was computed. To reduce sensitivity to tracking jitter, a circular tolerance region equal to 5\% of the bounding box diagonal was applied around the centroid. A detection was considered inside S or F if the buffered centroid intersected the corresponding rectangular zone. Directional classification was based on ordered zone transitions for each tracked individual (local\_ID): trajectories appearing first in S and then in F were classified as \textit{entry}, while those appearing first in F and later in S were classified as \textit{exit}. Trajectories exceeding a crossing duration of 10 seconds were labeled as uncertain and excluded.
Figure~\ref{fig:daily_entry_exit} shows the resulting daily entry and exit counts for Cameras 2 and 3. Bidirectional flow increases markedly during the May~2--4 festival window relative to off-peak days. Entry and exit counts remain largely symmetric across most days, indicating balanced inflow–outflow dynamics, while event periods reflect higher throughput rather than sustained accumulation.

\subsection{Movement Patterns}

Movement patterns can reveal customer habits, and responses to the environment. Typical approaches to tracking individual movement patterns using security cameras involve gait analysis, facial recognition, and color analysis. These additional visual markers can make it easy to connect paths that a computer vision algorithm might miss. Misidentification of individuals is common when two individuals cross paths. While such gaps are easily resolved in vision-rich systems through visual re-identification, ambiguity increases when looking at only the detection data. 

In addition to person misidentification, the trajectory data exhibit significant noise. Human movement is variable inherently, and trajectory analysis can be affected by incomplete paths and partial occlusions. Environmental factors further compound this issue and introduce new pattern variability. Two complementary analytical approaches were designed to extract actionable insights for owners. 

In both approaches, the first step is to patch the most obvious breaks in trajectory by attempting to reclassify misidentified individuals. Each point was put through an algorithm that processed the first and last point of each trajectory, finding trajectories that start near the same time, and merging these IDs if they are within a reasonable distance of each other. 

\subsubsection{Zoning Movement Approach}
\begin{figure}
    \centering
    \includegraphics[width=\linewidth]{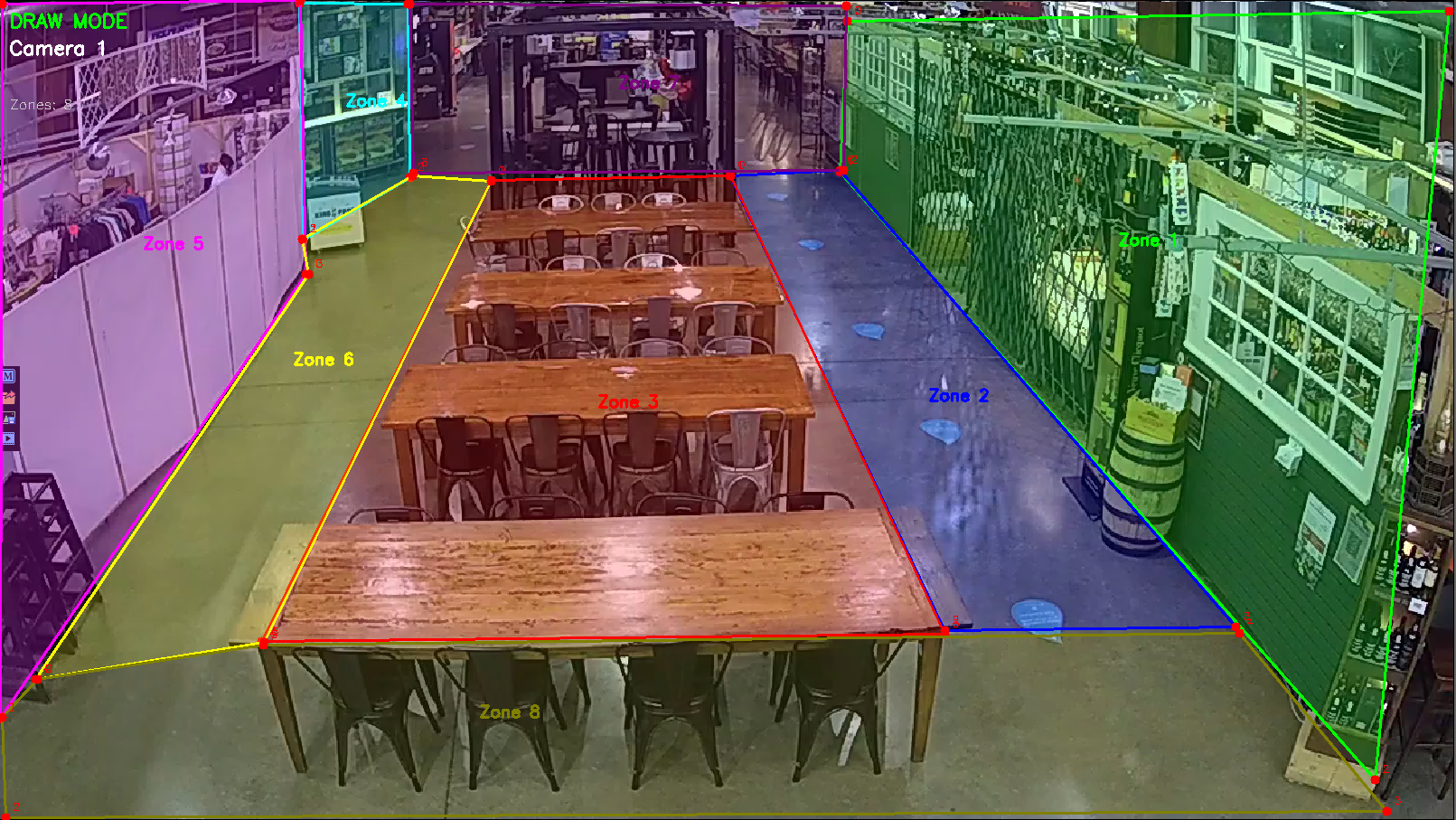}
    \caption{Zones of interest visible from Camera 1. The foreground and background are parted into their own zones, as well as the 3 main storefronts in Camera 1’s field of vision, Zone 1, 5 and 4}
    \label{fig:camera1zones}
\end{figure}

The first approach uses spatial zoning to survey movement data. The camera’s field of vision is separated into different zones of interest, such as entrances, aisles, seating areas and storefronts. Individual trajectories are mapped to sequences of zone transitions, rather than raw coordinate information. This representation circumvents the need for rigorous noise filtering, as the transition from zone to zone is the main concern. Figure \ref{fig:camera1zones} shows all ZOIs for Camera 1. Visually, the approach produces clear flow diagrams, making extracting meaningful spatial information simple. High-traffic transitions, customer defined zones of interest and probable subsequent zone are highlighted through this method.

\begin{figure}
    \centering
    \includegraphics[width=\linewidth]{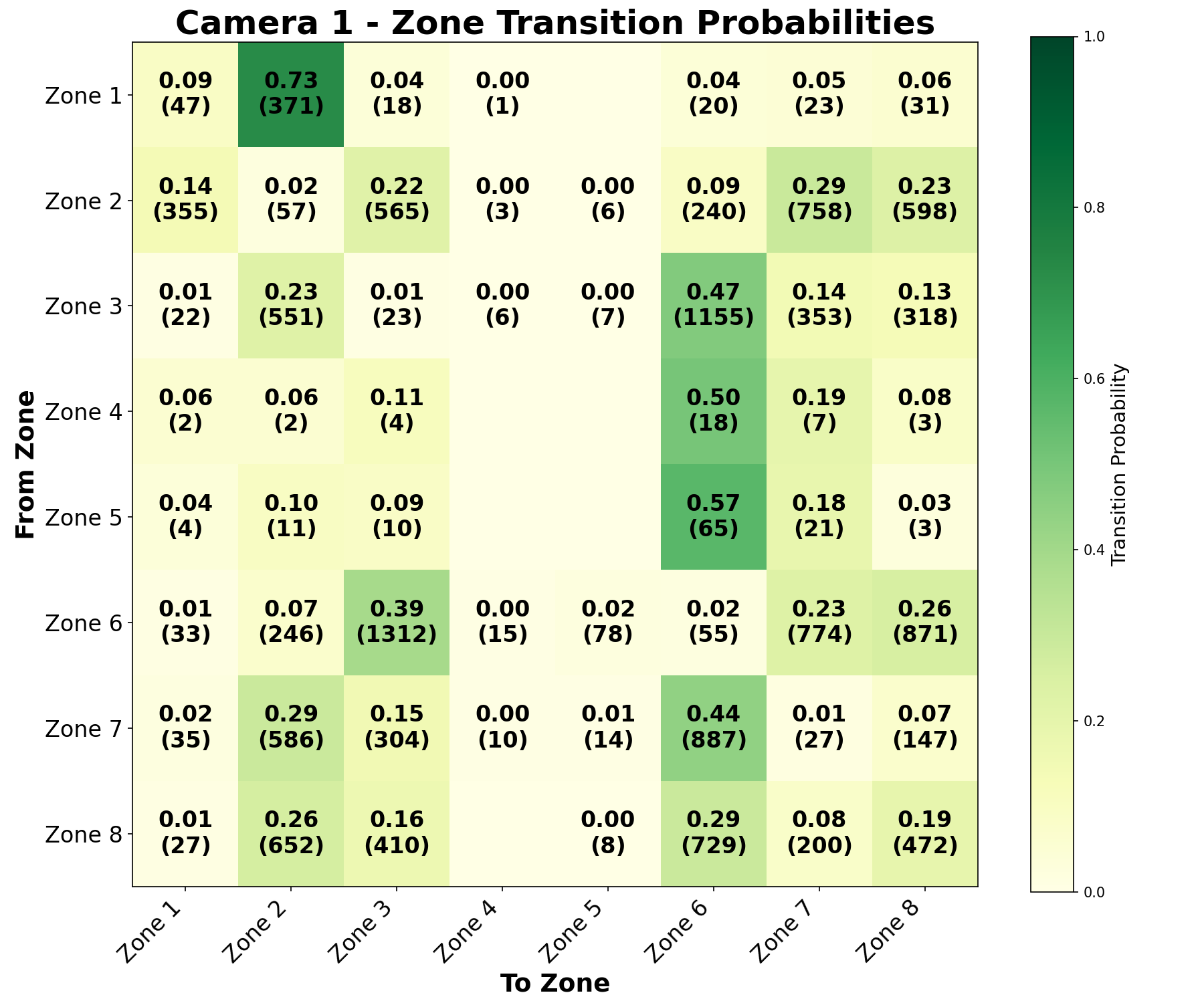}
    \caption{Figure shows the most probably next zone from the one of origin.}
    \label{fig:cam1transitionmatrix}
\end{figure}

Figure \ref{fig:cam1transitionmatrix} is the probability of customers moving throughout the zones found in Figure \ref{fig:camera1zones}. The zones of the most interest are Zones 2 and 6, which are pathways that customers take to stores. Customers in Zone 6 are most likely to transition to Zone 3 (the seating area), Zone 8 and Zone 7. Thus we can conclude that customers in Zone 6 are not likely to visit stores in Zones 4 and 5. Conversely, Zone 2, a similar pathway between Zones 7 and 8, shows transitions to the same zones and an additional significant amount of traffic to the store in Zone 1. These data can be used to better analyze the performance of these stores and marketing measures that can be taken to improve foot traffic. 

\subsubsection{Coordinate Movement Approach}
The second approach operates directly on trajectory information and focuses on the similarity between movement paths. Given two trajectories, defining them as close and similar comes down to overall shape and trajectory point order, rather than just point-wise comparison. For the purposes of determining the similarity between trajectories, the Fréchet distances were employed, falling back to Hausdorff distance when Fréchet distances were computationally expensive \cite{b36}. Fréchet distance is defined as 

\begin{equation}
d_F^{\,d}(P,Q) =
\min_{\sigma \in \Sigma}
\; \max_{k = 1,\dots,|\sigma|}
\left\| p_{\sigma_1(k)} - q_{\sigma_2(k)} \right\|
\end{equation}

Where $(P = (p_1,\dots,p_n)), (Q = (q_1,\dots,q_m))$
{\small
\begin{equation}
c(i,j) =
\max \Big(
    \|p_i - q_j\|,\;
    \min \big\{
        c(i-1,j),\;
        c(i-1,j-1),\;
        c(i,j-1)
    \big\}
\Big)
\end{equation}
}
with boundary conditions

\begin{equation}
\begin{aligned}
c(1,1) &= \|p_1 - q_1\|, \\
c(i,1) &= \max \big( \|p_i - q_1\|,\; c(i-1,1) \big), \\
c(1,j) &= \max \big( \|p_1 - q_j\|,\; c(1,j-1) \big).
\end{aligned}
\end{equation}

Fréchet distance is preferred due to its ability to respect path continuity and  traversal order. Each trajectory was resampled to 20 points. The pairwise distances between trajectories within the camera’s field of view were computed and subsequently clustered using DBSCAN. This density-based clustering approach was chosen due to its robust noise filtration. 
\begin{figure}
    \centering
    \includegraphics[width=\columnwidth]{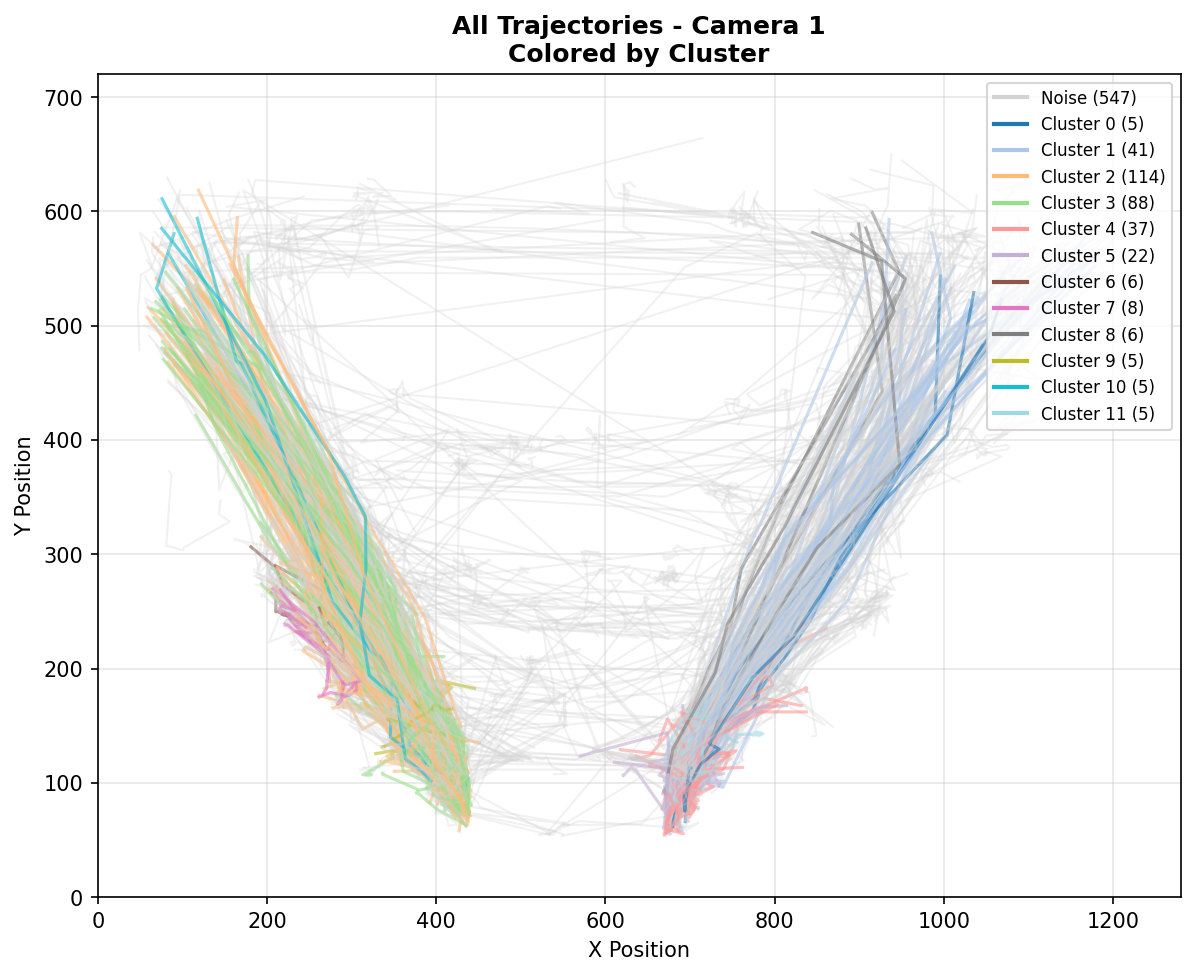}
    \caption{Most Popular Trajectories under Camera 1}
    \label{fig:camera1_traj_clusters}
\end{figure}

Figure \ref{fig:camera1_traj_clusters} shows the most popular trajectories in Camera 1. These data further illuminate the popularity of these pathways as thoroughfare but also highlight circular activity in Zone 8. The storefront beyond Zone 8 could use this insight to improve queue management, and the marketplace could attempt to direct waiting customers to other storefronts. 

\begin{figure}
    \centering
    \includegraphics[width=\columnwidth]{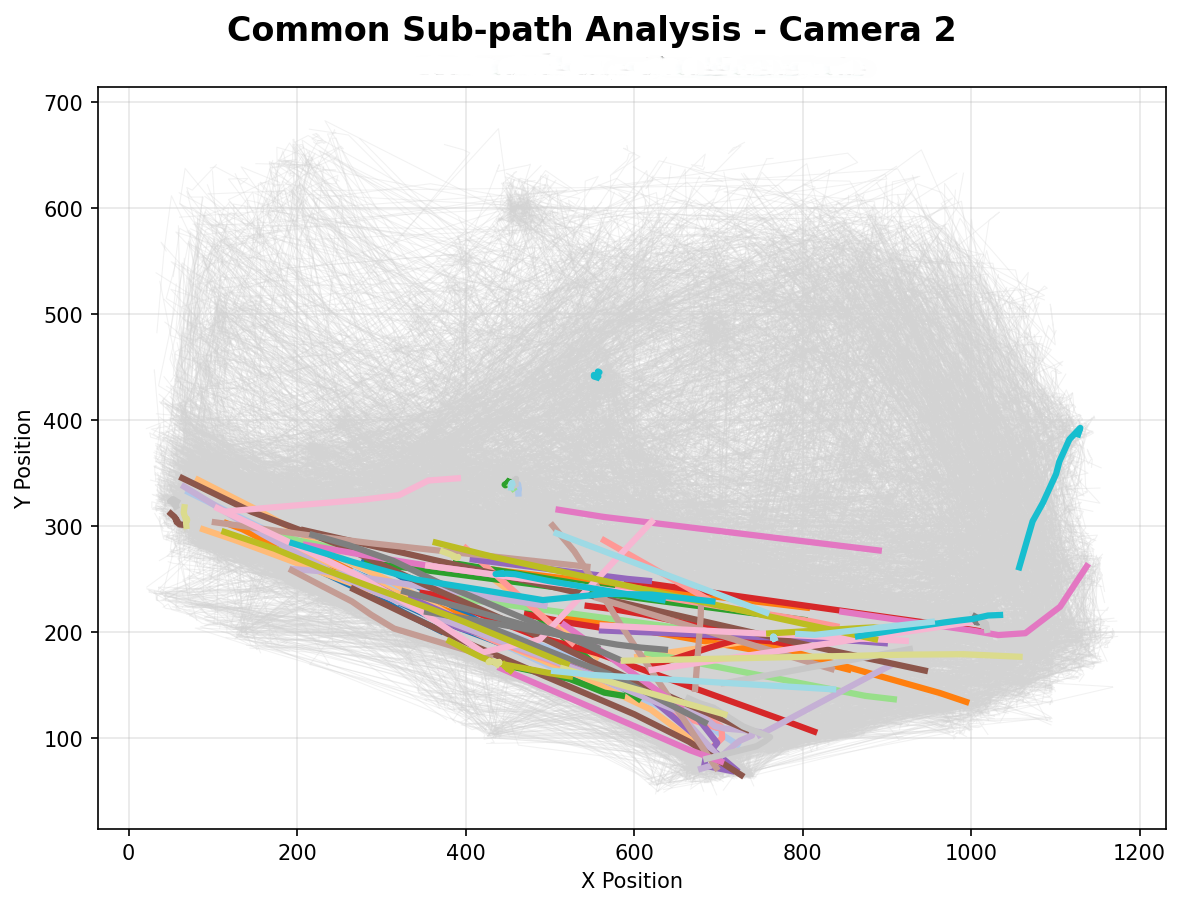}
    \caption{Most Popular Partial Trajectories under Camera 2}
    \label{fig:camera2_partial_traj}
\end{figure}

Further trajectory analysis found that there were sections of path similarities overlooked as a result of later path divergences. To address this, a segmentation approach was taken, where each trajectories 20 points were split into 8 point segments, with a 2 point sliding window of overlap. The similarity between these segments was analyzed, allowing for a more generalizable path comparison. The results of the path segmentation can be seen in Figure \ref{fig:camera2_partial_traj}. Where previous analysis showed only direct movement from the entryway to the storefront on the left, sub-path analysis clearly shows this is a widely traversed path that could be changed to avoid a bottleneck. 

\section{Discussion} 

In this study, we analyzed the temporal and spatial patterns of visitor behaviors to see whether these patterns are structured or random and the findings showed that metrics like dwell time and movement patterns are not only random but also dynamic and event-sensitive.
As shown in Figures \ref{fig:ridge_by_day} and Figure \ref{fig: daily_entry_exit}, both arrivals and dwell time increase on event days. In addition, Figure \ref{fig:camera1_traj_clusters} and Figure\ref{fig:camera2_partial_traj} show that movement paths are concentrated in specific areas, meaning that visitor traffic is not evenly distributed across the space.
These observed patterns are aligned with the consumer behavior literature, which demonstrates that purchasing is not a static act but a dynamic and multi-stage process shaped by the interaction between customer goals and the environment \cite{b28}, and that shopping behavior is path-dependent and can influence purchase probability and basket composition \cite{b29}. Studies also show that customers rarely visit the entire store and follow structured and uneven paths \cite{b30}. In addition, layout, volume, visual signals, and environmental design significantly affect time spent, search behavior, unplanned purchases, and brand evaluation \cite{b31} \cite{b32} and even longer in-store exposure increases the likelihood of impulse purchases \cite{b33} and layout and promotional placement influence sales behavior \cite{b32}. So measures such as dwell time, path tracking, movement between zones, and visual maps help reveal patterns of attention, stopping behavior, and decision steps.
In hospitality contexts, time spent at tables affects turnover, capacity, and revenue, making dwell time an operational variable, not just a behavioral one \cite{b34}.
So, while simple measures such as Traffic or Purchase Count are useful for reporting performance, they are not sufficient for optimizing layout, staffing, campaign design, or evaluating spatial fairness, and understanding these metrics can help business owners better understand consumer behavior. More detailed spatial–temporal metrics support deeper behavioral analysis and stronger operational and strategic decision-making. 

In Figure \ref{fig: daily_entry_exit}  we can see the total number of entris to the public market as a multiple vendor food hall with a shared sitting area at 2025-05-04 is about  600 people and in Figure \ref{fig:ridge_by_day}  we can see the dwell time is increasing at the same date. Table 1 shows mean dwell time
of 1528.3 seconds (˜25.5 minutes) on 2025-05-04, compared to 931.6 seconds (˜15.5 minutes) on a regular day like 2025-05-01. 

For a public market, when average dwell exceeds 20 minutes, management should consider temporary seating expansion, standing areas, targeted cleaning and table cleaning in 
high-dwell zones, and small layout or do readjustments to reduce congestion

Using time-varying demand logic \cite{b35}, system load can be approximated by the product of arrival rate and average service time. Approximating using Total counts, the operational load on 2025-05-04 relative to 2025-05-01 is about 6.6×. This shows that staffing based only on traffic or daily averages will underestimate peak load. Staffing should therefore be adjusted using a load multiplier, not only a traffic multiplier.

Figure \ref{fig:cam1transitionmatrix} shows that movement between zones is uneven. The probability of moving from Zone 1 to Zone 3 is about 0.73, while for some peripheral zones it is below 0.15, indicating about a 4.8× difference in exposure. This means vendor visibility is structurally unequal. An Exposure Index (sum of incoming transition probabilities) can identify high- and low-exposure zones and support rental differentiation or targeted promotional support.

Future research can merge these metrics with transactional sales data to see how changes in dynamic metrics like dwell time affects on revenue stream. It can also test in different locations or just during busy days to see how they can improve flow and performance.

\section*{Acknowledgment}
This research is supported by the National Science Foundation (NSF) under Award Number 2527312.

\end{document}